\pdfoutput=1
\documentclass[aps,pra,twocolumn,superscriptaddress,showpacs,preprintnumbers]{revtex4-1}
\usepackage[colorlinks=true, pdfstartview=FitV, linkcolor=red, citecolor=blue, urlcolor=black, pdftitle={},pdfauthor={Tomoya Hayata},pdfsubject={}, pdfkeywords={}]{hyperref}

\usepackage[dvipdfmx]{graphicx}
\usepackage{times}
\usepackage{amssymb,bm}
\usepackage{amsthm,amsmath}
\usepackage{color}

\newcommand{\cL}{{\cal L}}

\newcommand{\C}{{\rm C}}
\newcommand{\R}{{\rm R}}
\newcommand{\I}{{\rm I}}
\newcommand{\re}{{\rm Re}}
\newcommand{\im}{{\rm Im}}
\newcommand{\x}{\bm{x}}
\newcommand{\y}{\bm{y}}
\newcommand{\hi}{\hat{i}}
\newcommand{\vphi}{\varphi}

\begin{document}
\preprint{RIKEN-QHP-170}
\title{Complex Langevin simulation of quantum vortices in a Bose-Einstein condensate}
\author{Tomoya Hayata}
\affiliation{Department of Physics, The University of Tokyo, Tokyo 113-0031, Japan}
\affiliation{Theoretical Research Division, Nishina Center, RIKEN, Wako 351-0198, Japan}
\author{Arata Yamamoto}
\affiliation{Department of Physics, The University of Tokyo, Tokyo 113-0031, Japan}
\affiliation{Theoretical Research Division, Nishina Center, RIKEN, Wako 351-0198, Japan}

\date{\today}

\begin{abstract}
The ab-initio simulation of quantum vortices in a Bose-Einstein condensate is performed by adopting the complex Langevin techniques.
We simulate the nonrelativistic boson field theory at finite chemical potential under rotation.
In the superfluid phase, vortices are generated above  a critical angular velocity 
and the circulation is clearly quantized even in the presence of quantum fluctuations.
\end{abstract}

\maketitle

\section{Introduction}

Bose-Einstein condensation attracts a great deal of attention in various areas of physics.  
Direct observation has been achieved in experiments on liquid helium \cite{bec_He} and weakly interacting atomic gases~\cite{bec_coldatom,review}.
In solid state physics, the superconductivity of metals results from the condensation of Cooper pairs, which is a bound state of electrons in momentum space~\cite{Bardeen:1957mv}.
In the core of neutron stars, the condensate of the Cooper pairs of nucleons or quarks is considered to exist~\cite{Migdal:1978az}. 
Also, the condensation of the Higgs boson results in the dynamical mass generation of gauge bosons in the standard model of particle physics \cite{Englert:1964et}.

In the presence of external gauge fields, the Bose-Einstein condensate exhibits topological solitons.
In type-II superconductors under magnetic fields, the penetrating magnetic flux is quantized and the quanta form the Abrikosov lattice structure~\cite{Abrikosov:1956sx,sc:experiment}.
As understood from the analogy between magnetism and rotation, the quantum vortex has been observed in the rotating Bose-Einstein condensate \cite{experiment}.
It has been investigated in detail both from theories and experiments~\cite{Fetter:2009zz}. 

In dilute and low temperature systems, quantum and thermal fluctuations can be negligible, and thus the mean-field theory works well.
Quantum vortex nucleation in the Bose-Einstein condensate can be described by using the Gross-Pitaevskii equation~\cite{GP}.
However, when the quantum or thermal fluctuation becomes large, it is highly nontrivial how such topological solitons behave.
Around the critical values of temperature, chemical potential, magnetic field, or angular velocity, the fluctuation grows and then the mean-field description inevitably breaks down.
In fact, the effects of quantum fluctuations have been discussed in the literature~\cite{fluctuation}. 
Since high precision measurements are possible in cold-atom experiments,
the deviation from the mean-field theory can be detectable in experiments.
For a definite theoretical prediction without uncertainty, the ab-initio simulation of quantum vortex nucleation is necessary as shown in this paper.

In this paper, we report an ab-initio simulation of quantum vortices in the rotating Bose-Einstein condensate.
For this purpose, we adopt the complex Langevin method to nonrelativistic boson field theory.
The complex Langevin method has been developed in relativistic field theories to attack complex action problems, such as a nonequilibrium system~\cite{Berges:2005yt} and the phase diagram at finite quark number density~\cite{Karsch:1985cb}.
We first discuss the superfluid transition without rotation, and then analyze the vortex nucleation in the rotating frame.
We show that although the circulation is quantized in the superfluid phase,
quantum fluctuations blur the quantized circulation as a condensate fraction getting close to zero.

\section{Bose gas under rotation}

We consider quantum field theory at finite temperature, i.e., in ($1+3$)-dimensional Euclid spacetime.
In Euclid simulations, although we cannot follow real-time dynamics of vortex nucleation, which can be studied in the real-time Gross-Pitaevskii simulation~\cite{Tsubota}, 
we can still study the nonperturbative mechanism of it. 

The continuum action of a complex boson field $\vphi(\tau,\x) = \vphi^1(\tau,\x) + i\vphi^2(\tau,\x)$ in a rotating frame is~\cite{Andersen:2004} 
\begin{equation}
\begin{split}
& S_{\rm con} [\vphi^1,\vphi^2] 
\\
=& \int d\tau d^3x\;\Bigl[ \vphi^{*}(\partial_\tau-\mu)\vphi+\frac{1}{2m}|(\bm{\nabla}-im\bm{\Omega}\times\x)\vphi|^2 
\\
& -\frac{1}{2}m(x^2+y^2)\Omega^2|\vphi|^2+\frac{1}{4}\lambda |\vphi|^4
\Bigr], 
\end{split}
\label{action:continuum}
\end{equation}
where $\mu$, $m$ and $\bm{\nabla}$ denote the chemical potential, mass of the boson, and spatial derivatives, respectively.
We consider the rotation around the $z$ axis with angular velocity $\Omega$, and thus $\bm{\Omega}= \Omega \hat{z}$.
($\hi$ denotes a unit vector in the $i$ direction.)

We remark here that, except for the centrifugal potential $-\frac{1}{2}m(x^2+y^2)\Omega^2 |\vphi|^2$, 
the action~\eqref{action:continuum} is mathematically equivalent to the spinless charged boson action under a magnetic field. 
In the rotating frame, particles effectively couple to the ``magnetic field'', 
\begin{equation}
q\bm{B}=q \bm{\nabla} \times \bm{A}=2m\Omega \hat{z},
\end{equation}
 with $\bm{A}=\bm{\Omega}\times\x$.
 Therefore, our analysis can be applied not only to the rotation but also to the magnetic field.
In fact, the qualitative behavior of vortex nucleation is the same in the case of a magnetic field.
In the following, we only show the result of rotation.

To perform a lattice simulation, we discretize the continuum action~\eqref{action:continuum} on the hypercubic lattice.
The corresponding lattice action is
\begin{equation}
\begin{split}
& S_{\rm lat} [\vphi^1,\vphi^2]
\\
&= a^3\sum_{\tau,\x}\;\Bigl[ \vphi^{*}_{\tau,\x}\left(\vphi_{\tau,\x}-e^{\mu a}\vphi_{\tau-a,\x}\right) 
\\
&-\frac{1}{2ma}\sum_i\left(\vphi^{*}_{\tau,\x+\hi a}u^{\dagger}_i\vphi_{\tau,\x}+\vphi^{*}_{\tau,\x}u_i\vphi_{\tau,\x+\hi a}-2|\vphi_{\tau,\x}|^2\right)
\\
& -\frac{1}{2}ma(x^2+y^2)\Omega^2\vphi^{*}_{\tau,\x}\vphi_{\tau-a,\x}
+\frac{1}{4}\frac{\lambda}{a^2} a^3\left(\vphi^{*}_{\tau,\x}\vphi_{\tau-a,\x}\right)^2 \Bigr],
\end{split}
\label{action:lattice}
\end{equation}
where $a$ is the lattice spacing.
The effective gauge field of rotation is introduced in the same manner as the electromagnetic gauge field~\cite{D'Elia:2012tr},
\begin{equation}
 u_i = \exp(-ia q A_i ) = \exp(-iam (\bm{\Omega}\times\x)_i ) .
\end{equation}
The chemical potential is introduced on the basis of the standard lattice formulation~\cite{Hasenfratz:1983ba}.
The centrifugal potential and the contact interaction terms are regularized in the same manner as the number density operator.

In the path integral quantization, the expectation value of the operator $\hat{O}$ is given by
\begin{eqnarray}
&&\langle \hat{O} \rangle=  \frac{1}{Z}\int d\vphi^1 d\vphi^2 \;e^{-S_{\rm lat}[\vphi^1,\vphi^2]} \hat{O} [\vphi^1,\vphi^2] ,
\label{eq:expectation}
\end{eqnarray}
where $Z$ is a normalization factor, 
\begin{equation}
Z=\int d\vphi^1 d\vphi^2 \ e^{-S_{\rm lat}} .
\end{equation}
In the conventional quantum Monte Carlo simulations based on importance sampling techniques, one evaluates Eq.~\eqref{eq:expectation} 
by means of the ensemble average, which is randomly generated by the probability density $e^{-S_{\rm lat}}/ Z$. 
However, as discussed in Ref.~\cite{Yamamoto:2012wy}, the lattice action~\eqref{action:lattice} suffers from the notorious sign problem
because the temporal hopping term is in general complex in nonrelativistic systems. 
The probability interpretation of the weight $e^{-S_{\rm lat}}/ Z$ breaks down and thus the importance sampling cannot be applied. 
To overcome this difficulty, we adopt the complex Langevin technique, 
which is based on the stochastic quantization formalism 
and does not necessarily require the action to be real.

\section{Complex Langevin method}
In this method, we numerically perform the stochastic quantization for the complex lattice action~\eqref{action:lattice}.
The stochastic quantization reconstructs Eq.~\eqref{eq:expectation} 
by the noise average of the solution of classical equation of motion with random noises~\cite{Parisi:1980ys,Damgaard:1987rr}.
For this purpose, we need to solve the Langevin equations for $\vphi^a$ ($a=1,2$) along the fictitious time direction 
\begin{eqnarray}
\partial_\theta \vphi_{\tau,\x}^a(\theta)= -\frac{\partial S_{\rm lat}[\vphi^1,\vphi^2] }{\partial \vphi^a_{\tau,\x}} +\eta^a_{\tau,\x}(\theta),
\label{eq:Langevin}
\end{eqnarray}
with $\theta$ being the continuous fictitious time, 
and $\eta^a_{\tau,\x}(\theta)$ real Gaussian noises.
Since the lattice action~\eqref{action:lattice} is complex, the right-hand side of Eq.~\eqref{eq:Langevin} is complex.
Thus, we need to complexify the left-hand side, i.e., complexify the two real fields as $\vphi^a \to \vphi^{a\C}=\vphi^{a\R}+i\vphi^{a\I}$. 
Then, Eq.~\eqref{eq:Langevin} becomes a stochastic differential equation for the two complex fields $\vphi^{a\C}$ ($a=1,2$),
in which the Gaussian noises are applied only to the real parts~\cite{Parisi:1980ys,Damgaard:1987rr}.
Now, Eq.~\eqref{eq:Langevin} reads
\begin{eqnarray}
\partial_\theta \vphi_{\tau,\x}^{a \R} (\theta) &=& -\re\left[\frac{\partial S_{\rm lat}[\vphi^{1\C},\vphi^{2\C}] }{\partial \vphi_{\tau,\x}^{a\C}}\right] +\eta^a_{\tau,\x}(\theta),
\label{eq:Langevin_re} \\
\partial_\theta \vphi_{\tau,\x}^{a \I} (\theta) &=& -\im\left[\frac{\partial S_{\rm lat}[\vphi^{1\C},\vphi^{2\C}]}{\partial \vphi_{\tau,\x}^{a\C}}\right] .
\label{eq:Langevin_im}
\end{eqnarray}
The Gaussian noises $\eta^a_{\tau,\x}(\theta)$ satisfy 
\begin{eqnarray}
\langle \eta^a_{\tau,\x}(\theta)\rangle_\eta &=& 0 , \\
\langle \eta^a_{\tau,\x}(\theta)\eta^b_{\tau^\prime, \x^\prime}(\theta^\prime) \rangle_\eta &=& 2\delta_{ab}\delta_{\tau\tau^{\prime}}\delta_{\x\x^\prime}\delta(\theta-\theta^\prime).
\end{eqnarray}
We note that, since the order parameter is the field variable itself, the mean-field calculation is equivalent to the classical calculation in this particular theory.
The classical solution is given by minimizing the action~\eqref{action:continuum} or~\eqref{action:lattice}, 
which is nothing but the equilibrium solution of Eqs.~\eqref{eq:Langevin_re} and~\eqref{eq:Langevin_im} without the random noises.

The expectation value~\eqref{eq:expectation} is obtained from the solution of Eqs.~\eqref{eq:Langevin_re} and~\eqref{eq:Langevin_im} as
\begin{eqnarray}
&&\langle \hat{O} \rangle=  \lim_{\theta\to\infty}\langle \hat{O} [\vphi^{1\C}(\theta),\vphi^{2\C}(\theta)] \rangle_\eta ,
\label{eq:expectation_langevin}
\end{eqnarray}
where the operator is written in terms of the complex fields $\vphi^{a\C}$.
For example, the number density operator, $\hat{n}_{\tau,\x}=-\partial \cL/\partial \mu$ with $\cL$ being the lattice Lagrangian density, reads 
\begin{eqnarray}
 \hat{n}_{\tau,\x} &=&   e^{\mu a}\left(\delta_{ab}+i\epsilon_{ab}\right)\vphi_{\tau,\x}^{a\C}\vphi_{\tau-a,\x}^{b\C} ,
\label{eq:density}
\end{eqnarray}
where $\epsilon_{ab}$ is a completely antisymmetric tensor with $\epsilon_{01}=1$ and the Einstein convention is understood for repeated indices.
Other observables are complexified in the same manner and have both real and imaginary parts.

\section{Numerical simulation}

We numerically solved Eqs.~\eqref{eq:Langevin_re} and~\eqref{eq:Langevin_im} 
with the fictitious time step $\varepsilon=1.0\times10^{-4}a$.
We adopted a higher order algorithm used in Ref.~\cite{Aarts:2011zn} to improve the step size dependence~\cite{chang}. 
Errors were estimated by using the jackknife method.
Although the complex Langevin simulation is known to be unreliable in some examples at hand~\cite{Aarts:2013uxa}, we found no undesirable or pathological behavior in our simulations.
A singular drift term in Eqs.~\eqref{eq:Langevin_re} or~\eqref{eq:Langevin_im}, which leads to the failure of the complex Langevin simulation, does not appear in our model~\cite{Nishimura:2015pba}.

\begin{figure}[h]
\centering
\includegraphics[scale=.83]{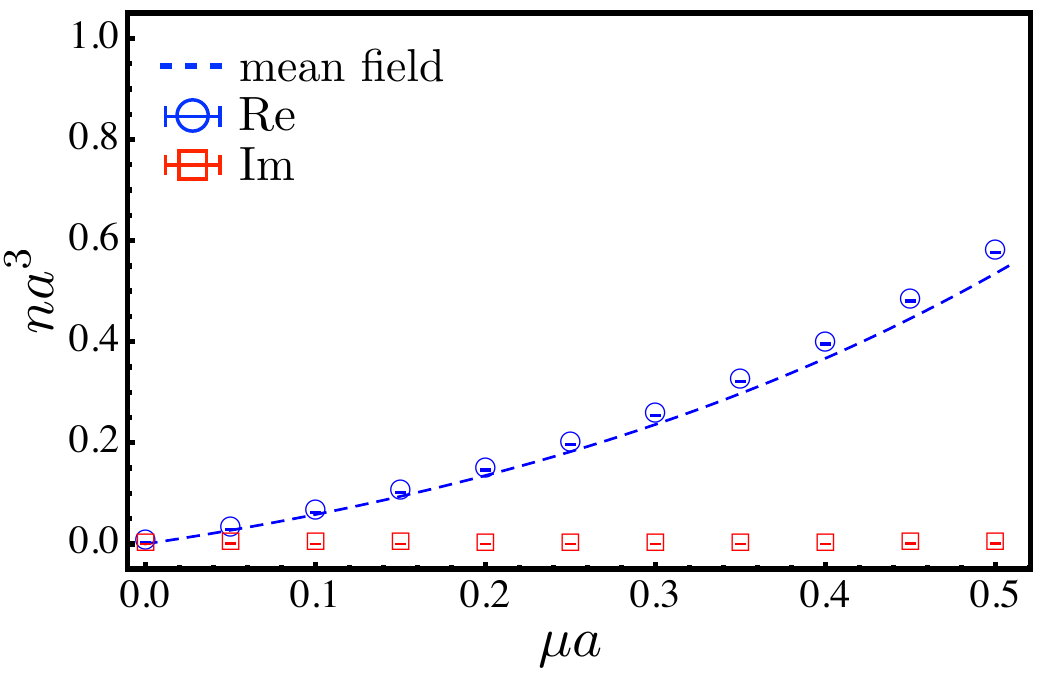}
\caption{\label{fig1}
(Color online) Number density $n$ in periodic boundary conditions at $\Omega =0$.
}
\end{figure}
\begin{figure}[h]
\centering
\includegraphics[scale=.83]{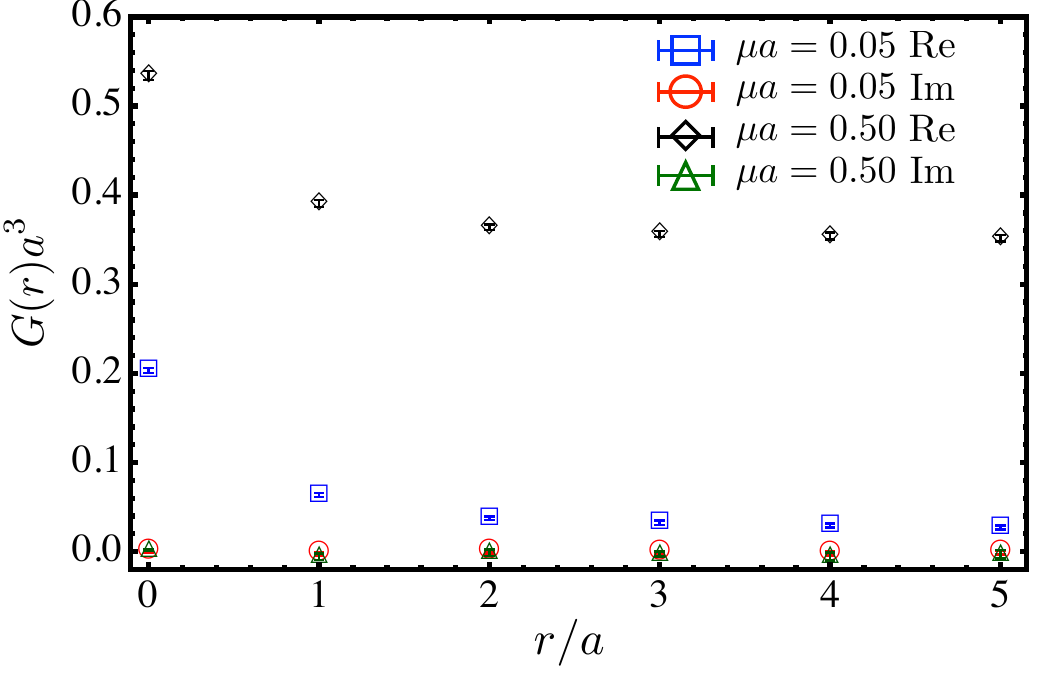}%
\caption{\label{fig2}
(Color online) Two-point correlation function $G(r)$ at $\mu a=0.05$ and $0.50$.
}%
\end{figure}%
\begin{figure}[h]
\centering
\includegraphics[scale=.83]{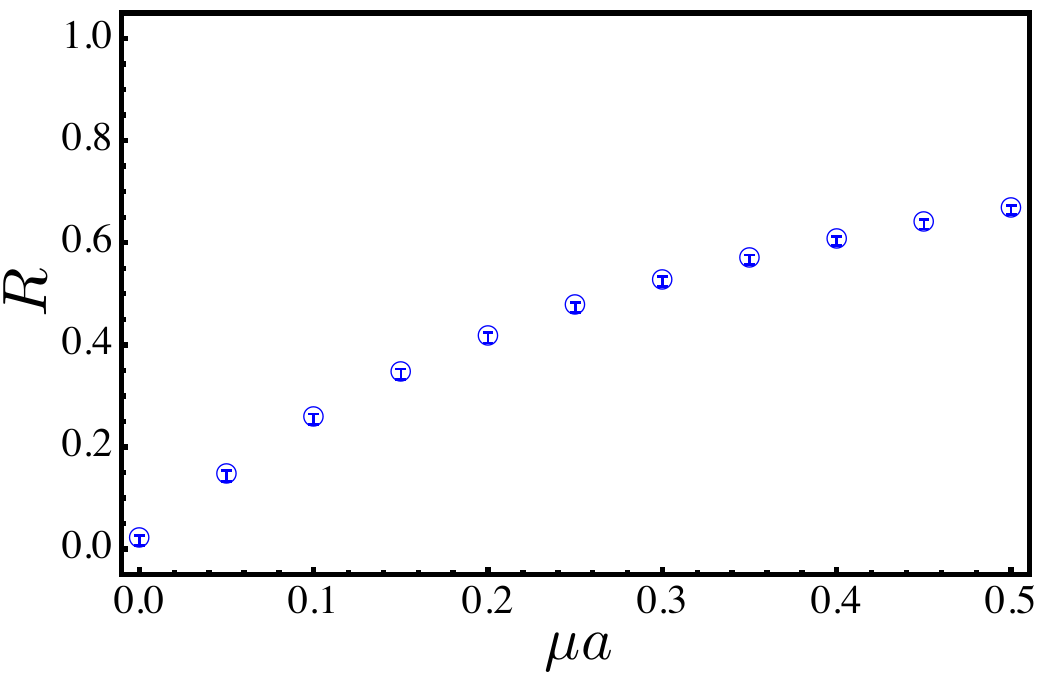}%
\caption{\label{fig3}
(Color online) Condensate fraction $R$ in periodic boundary conditions at $\Omega =0$.
}%
\end{figure}%
\begin{figure}[h]
\centering
\includegraphics[scale=.82]{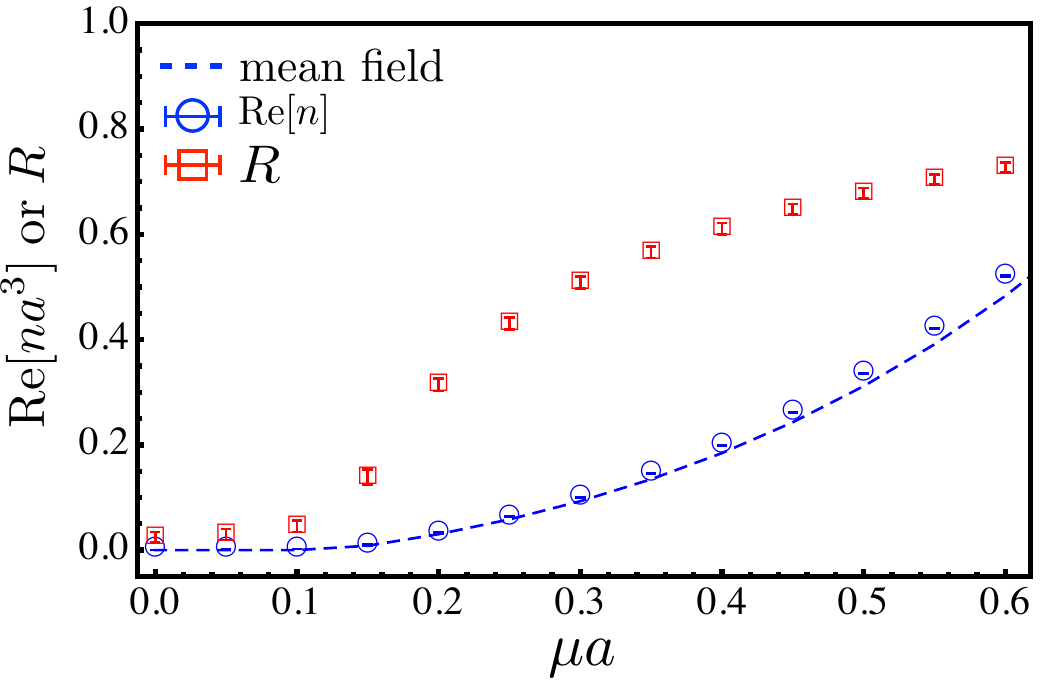}%
\caption{\label{fig4}
(Color online) Condensate fraction $R$ and number density $n$ in the Dirichlet boundary conditions at $\Omega =0$.
}%
\end{figure}%

First, we analyze the superfluid transition without rotation, i.e., $\Omega=0$.
The total number of lattice sites is $V=N_xN_yN_zN_\tau=12^4$.
Here we take the periodic boundary conditions in all directions.
We set $ma=0.50$ and $\lambda/a^2=4.0$.

We show the number density 
\begin{equation}
 n = \frac{1}{V} \sum_{\tau,\x} \langle \hat{n}_{\tau,\x} \rangle
\end{equation}
as a function of a chemical potential in Fig.~\ref{fig1}.
In the nonrelativistic Bose gas, the number density increases without a gap as the chemical potential increases,
thus there is no ``Silver Blaze'' problem~\cite{Cohen:2003kd}.
In periodic boundary conditions, we can analytically minimize the lattice action~\eqref{action:lattice} at $\Omega a=0$, 
and obtain the gapless mean-field solution  $n = 2e^{\mu a}(e^{\mu a}-1)/(\lambda a)$, which is also shown in Fig.~\ref{fig1}.
Our calculation is consistent with the mean-field solution at small chemical potentials but deviates from it at large chemical potentials.
The deviation from the mean-field solution seems opposite to that of the loop correction~\cite{Andersen:2004}.
We checked that the deviation is suppressed on a finer lattice and thus it is a lattice discretization artifact.
The continuum limit must be carefully taken to discuss the quantitative comparison with analytical results in a continuum space.

The two-point correlation function
\begin{equation}
 G(|\x-\y|)= \langle (\delta_{ab}+i\epsilon_{ab}) \vphi^{a\C}_{\tau,\x} \vphi^{b\C}_{\tau,\y} \rangle 
\end{equation}
at $\mu a=0.05$, and $0.50$ is shown  in Fig.~\ref{fig2}.
We see the signature of the off-diagonal long-range order~\cite{Yang1962} at $\mu a=0.50$.
From the correlation function, we define the condensate fraction
\begin{equation}
R = \frac{ {\rm Re} [G(aN_z/2)] }{ {\rm Re} [G(0)] },
\end{equation}
which is real by its definition.
The result is shown in Fig.~\ref{fig3}.
The browup of the condensate fraction cannot be obtained in the mean-field calculation because the mean-field condensate fraction is always unity at zero temperature.

As shown in Figs.~\ref{fig1} and \ref{fig2}, the imaginary parts are zero with sufficiently small error bars.
This is a requirement for the validity of the complex Langevin simulation.
The same is true and the imaginary parts are not shown in the figures below. 

Next, we change boundary conditions to simulate a rotating system. 
We take the Dirichlet boundary conditions in the $x$ and $y$ directions, and take periodic boundary conditions in the $z$ and $\tau$ directions.
We also change the lattice volume to $V=N_xN_y \times N_zN_\tau=11^2\times10^2$, 
where $x$ and $y$ are in the range $[-5a,5a]$ and the position of the rotational axis is set to $(x,y)=(0,0)$.
We have checked the volume independence of the following discussion by using a larger lattice volume,  $V=13^2\times12^2$.
Other parameters are the same as above.

In Fig.~\ref{fig4}, we show the condensate fraction $R$ and the number density $n$ as functions of a chemical potential $\mu$.
A nonzero gap exists
because of the inhomogeneity in the Dirichlet boundary conditions.
The transition to the superfluid phase can be clearly seen at $\mu a\sim 0.1$, which is accompanied by a browup of the condensate fraction.
Although this is the artificial transition in the Dirichlet boundary conditions, there is a physical phase transition at a higher temperature.
The critical exponents of these transitions can be estimated from power-law behaviors near the critical point by using the so-called finite-size scaling method.

\begin{figure}[h]
\centering
\includegraphics[scale=.83]{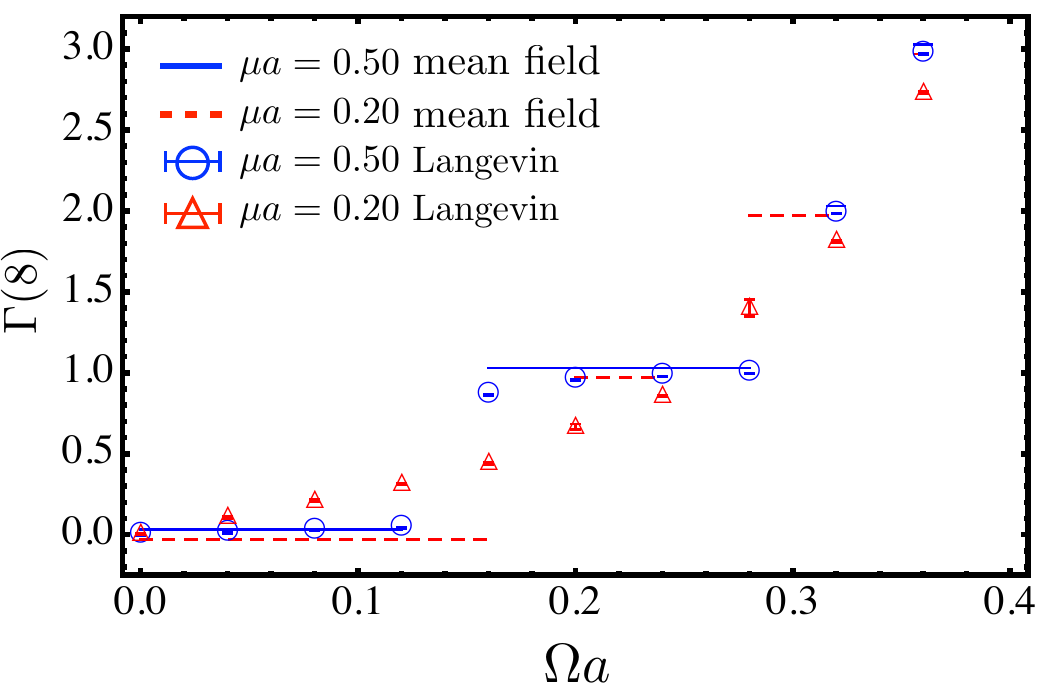}
\caption{\label{fig5}
(Color online) Circulation $\Gamma(8)$ as a function of angular velocity $\Omega$ at $\mu a=0.20$ and $0.50$.
}
\end{figure}
\begin{figure*}[ht]
\centering
\includegraphics[scale=.84]{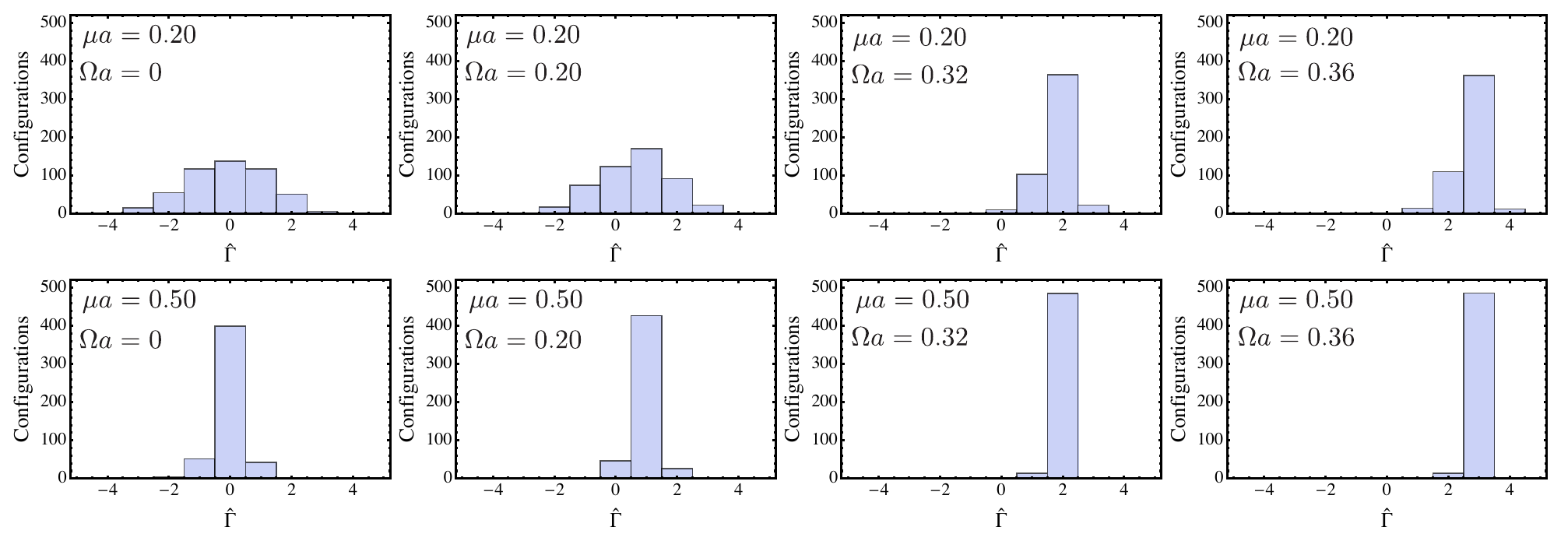}%
\caption{\label{fig6}
(Color online) Profile of the circulation $\hat{\Gamma}(8)$.
}%
\end{figure*}%

The direct evidence of a quantum vortex is the quantized circulation.
The circulation of a vortex is defined as the phase integral around it.
On the hypercubic lattice, the circulation is given by integrating the phase difference along the square loop.
We calculated the circulation of the U(1) phase of $\vphi^{\C}=\vphi^{1\C}+i\vphi^{2\C}$:
\begin{equation}
\begin{split}
& \hat{\Gamma}(l) 
\\
&= \oint_{l \times l} \frac{d\bm{x}}{2\pi} \bigg[ \tan^{-1} \bigg( \frac{\im[\vphi^{\C}_{\tau,\x+\bm{j}}]}{\re[\vphi^{\C}_{\tau,\x+\bm{j}}]} \bigg) - \tan^{-1} \bigg( \frac{\im[\vphi^{\C}_{\tau,\x}]}{\re[\vphi^{\C}_{\tau,\x}]} \bigg) \bigg] ,
\end{split}
\label{eq:circulation}
\end{equation}
where $\bm{j}$ is a unit vector along the loop.
The size of the loop is $l \times l$ ($2\leq l \leq10$ in our simulation), and the center of the loop is placed at $(x,y) = (0,0)$.
In each configuration of the ensemble, $\hat{\Gamma}(l)$ is an integer because of the single-valuedness of the wave functions, 
but it is not necessarily the same in different configurations. 
The ensemble average $\Gamma(l) \equiv \langle \hat{\Gamma}(l) \rangle$ becomes a non-integer if the number of vortices strongly fluctuates.  
The mean-field theory, in which the circulation takes an exact integer value, works well when the fluctuation is negligible.

In Fig.~\ref{fig5}, we show the circulation as a function of angular velocity 
in the superfluid phase with small ($\mu a=0.20$) and large ($\mu a=0.50$) condensate fractions.
At $\mu a=0.50$, the circulation is clearly quantized and consistent with the mean-field calculation.
On the other hand, at $\mu a=0.20$, it is not quantized and deviates from the mean-field prediction. 
This indicates the break down of the mean-field theory as the condensate fraction getting close to zero.
To see the fluctuation of vortices, we show the profile of circulation $\hat{\Gamma}$ obtained from each configuration at $\mu a=0.20$ and $0.50$ in Fig.~\ref{fig6}. 
At $\mu a=0.20$, the profile shows a Gaussian-like distribution.
The system is given by a superposition of different vortex numbers.
At $\mu a=0.50$, the profile becomes almost a single peak.
The fluctuation can be negligible and thus the mean-field theory works well as shown in Fig.~\ref{fig5}.
The change of the distribution suggests the change of the energy spectrum obtained by diagonalizing the Hamiltonian \cite{fluctuation}.
A Gaussian-like distribution suggests a small excitation energy to other vortex numbers and a single-peak distribution suggests a large excitation energy.

\begin{figure}[h]
\centering
\includegraphics[scale=.83]{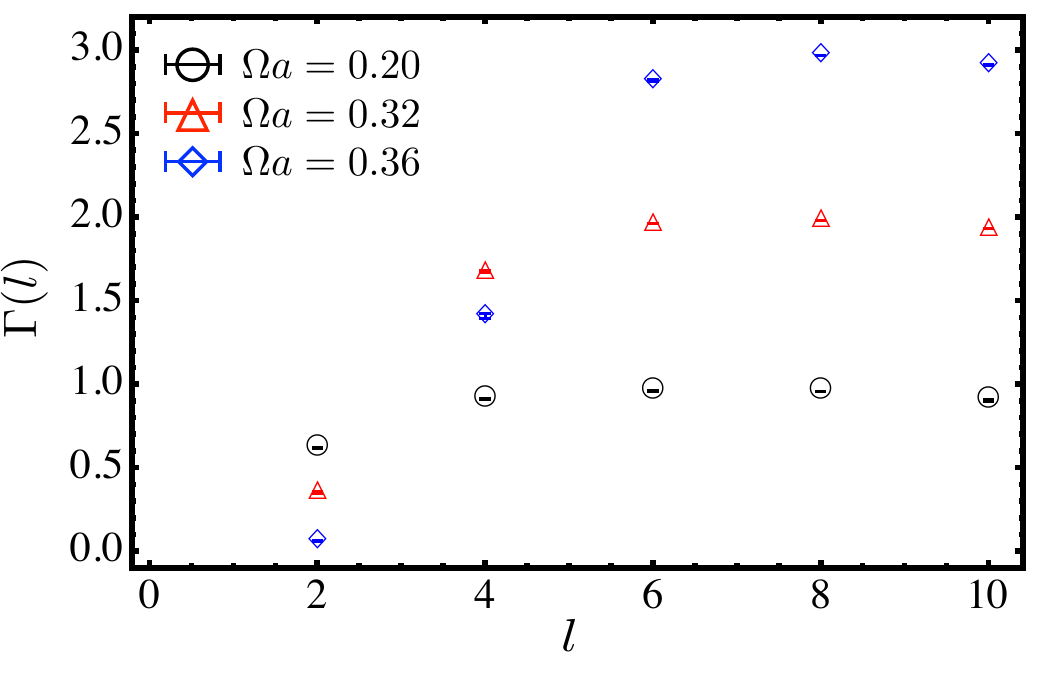}%
\caption{\label{fig7}
(Color online) Circulation $\Gamma(l)$ as a function of loop size $l$ at $\mu a=0.50$, and $\Omega a=0.20$, $0.32$ and $0.36$.
}%
\end{figure}

The spatial positions of vortices can be estimated from the $l$-dependence of $\Gamma(l)$ shown in Fig.~\ref{fig7}.
At $\Omega a=0.20$ (circles), $\Gamma(l)=1$ and it is almost independent of $l$ in $l\ge 2$.
Thus, one vortex exists inside the $2 \times 2$ loop, i.e., in $|x| \le a$ and $|y| \le a$.
At $\Omega a=0.32$ (triangles), $\Gamma(l)$ increases at $l=4$, and thus two vortices exist in $a \le |x| \le 2a$ and $a \le |y| \le 2a$.
At $\Omega a=0.36$ (diamonds), two vortices exist in $a \le |x| \le 2a$ and $a \le |y| \le 2a$ and one vortex exists $2a \le |x| \le 3a$ and $2a \le |y| \le 3a$.
By calculating two-point or three-point correlation functions of loops, we can obtain more detailed information, such as the intervortex distance or the Abrikosov lattice structure, although we must use a finer lattice than the present one.

\section{Concluding remarks}
We have performed an ab-initio simulation of quantum vortices in the Bose-Einstein condensate.
We have adopted the complex Langevin method instead of the quantum Monte Carlo method, 
which suffers from a sign problem because of the complex action~\eqref{action:lattice}.
In this ab-initio simulation, all quantum fluctuations are exactly taken into account. 
We have successfully simulated vortices in the rotating Bose-Einstein condensate.
We have shown that in the superfluid phase with a large condensate fraction, the fluctuation of vortices is negligible and the mean-field theory works well.
On the other hand, as the condensate fraction gets smaller, the fluctuation becomes larger, which shows a Gaussian-like distribution. 
The mean-field theory breaks down, and the circulation is not quantized even in the superfluid phase. 

There are several future applications.
We can calculate the two-body scattering length, which is commonly used to fix the physical scale to compare theoretical calculations with experiments, 
by the L\"uscher formula \cite{Luscher:1986pf}.
We can apply the anisotropic harmonic trapping potential, which corresponds to the time-dependent rotating potential in cold-atom experiments \cite{experiment}.
By tuning the trapping potential and angular velocity, we can study the quantum Hall effect expected in the rotating Bose-Einstein condensate \cite{Viefers}.
Our approach is also applicable to two dimensional Bose gases, where quantum fluctuation can be much stronger than that in three dimensions.
It is interesting to analyze quantum vortices in two dimensions
 near the quasi long-range order phase-transition point~\cite{BKT,BKTexperiments}.
Another application is relativistic field theory under rotation, where the sign problem exists~\cite{Yamamoto:2013zwa}.
Nonperturbative study of relativistic vortices is helpful in the understanding of the physics of, 
e.g., vortices inside neutron stars \cite{Migdal:1978az} and cosmic strings in the early universe~\cite{Kibble:1976sj}.

 The complex Langevin method is powerful not only for the rotating Bose-Einstein condensate but also for other condensed matter systems.
 First, nonperturbative simulations at finite temperatures are interesting. 
 We can estimate the critical temperature of the Bose-Einstein condensation, including all orders of density corrections~\cite{tc}.
 We can also study the effect of thermal fluctuation to the condensate and discuss the break down of the Gross-Pitaevskii equation.
 Such a break down of the Gross-Pitaevskii equation can be analyzed in cold atom experiments, where 
 a similar vortex fluctuation would be observed as the temperature gets close to its critical value.
\\

\begin{acknowledgements}
The authors thank T.~Doi, T.~Hatsuda, Y.~Hidaka and Y.~Tanizaki for stimulating discussions and helpful comments.
T.~H. was supported by JSPS Research Fellowships for Young Scientists.
\end{acknowledgements}

%------------------------------------------------------------


\begin{thebibliography}{100}

\bibitem{bec_He}
P.~Kapitza, 
Nature {\bf 141}, 74 (1938);
J.~F.~Allen and A.~D.~Misener,
Nature {\bf 142}, 643 (1938);
D.~D.~Osheroff, R.~C.~Richardson, and D.~M.~Lee,
Phys.\ Rev.\ Lett.\ {\bf 28}, 885 (1972).

\bibitem{bec_coldatom}
 M.~H.~Anderson, J.~R.~Ensher, M.~R.~Matthews, C.~E.~Wieman, and E.~A.~Cornell,
 Science {\bf 269} 198 (1995);
 C.~C.~Bradley, C.~A.~Sackett, J.~J.~Tollett, and R.~G.~Hulet,
 Phys.\ Rev.\ Lett.\ {\bf 75}, 1687 (1995);
 K.~B.~Davis, M.~-O.~Mewes, M.~R.~Andrews, N.~J.~van~Druten, D.~S.~Durfee, D.~M.~Kurn, and W.~Ketterle,
 Phys.\ Rev.\ Lett.\ {\bf 75}, 3969 (1995).

\bibitem{review}
 I.~Bloch, J.~Dalibard, and W.~Zwerger, Rev.\ Mod.\ Phys.\ {\bf 80}, 885 (2008).

\bibitem{Bardeen:1957mv} 
  J.~Bardeen, L.~N.~Cooper and J.~R.~Schrieffer,
  %``Theory of superconductivity,''
  Phys.\ Rev.\  {\bf 108}, 1175 (1957).

\bibitem{Migdal:1978az} 
  A.~B.~Migdal,
  %``Pion Fields in Nuclear Matter,''
  Rev.\ Mod.\ Phys.\  {\bf 50}, 107 (1978);
%\bibitem{Alford:2007xm} 
 M.~G.~Alford, A.~Schmitt, K.~Rajagopal and T.~Sch\"afer,
  %``Color superconductivity in dense quark matter,''
  Rev.\ Mod.\ Phys.\  {\bf 80}, 1455 (2008)
  [arXiv:0709.4635 [hep-ph]];
%\bibitem{Eto:2013hoa} 
  M.~Eto, Y.~Hirono, M.~Nitta and S.~Yasui,
  %``Vortices and Other Topological Solitons in Dense Quark Matter,''
  PTEP {\bf 2014}, 012D01 (2014)
  [arXiv:1308.1535 [hep-ph]].

\bibitem{Englert:1964et} 
  F.~Englert and R.~Brout,
  %``Broken Symmetry and the Mass of Gauge Vector Mesons,''
  Phys.\ Rev.\ Lett.\  {\bf 13}, 321 (1964);
  P.~W.~Higgs,
  %``Broken Symmetries and the Masses of Gauge Bosons,''
  Phys.\ Rev.\ Lett.\  {\bf 13}, 508 (1964).  

\bibitem{Abrikosov:1956sx} 
  A.~A.~Abrikosov,
  %``On the Magnetic properties of superconductors of the second group,''
  Sov.\ Phys.\ JETP {\bf 5}, 1174 (1957)
  [Zh.\ Eksp.\ Teor.\ Fiz.\  {\bf 32}, 1442 (1957)];
  W.~H.~Kleiner, L.~M.~Roth, and S.~H.~Autler,
  Phys.\ Rev.\ {\bf 133}, A1226 (1964).

\bibitem{sc:experiment}
U.~Essmann, H.~Tr{\"a}uble, 
Phys.\ Lett.\ A {\bf 24}, 526 (1967);
H.~F.~Hess, R.~B.~Robinson, R.~C.~Dynes, J.~M.~Valles,~Jr., and J.~V.~Waszczak,
Phys.\ Rev.\ Lett.\ {\bf 62}, 214 (1989).
  
\bibitem{experiment}
M.~R.~Matthews, B.~P.~Anderson, P.~C.~Haljan, D.~S.~Hall, C.~E.~Wieman, and E.~A.~Cornell,
Phys.\ Rev.\ Lett.\ {\bf 83}, 2498 (1999);
K.~W.~Madison, F.~Chevy, W.~Wohlleben, and J.~Dalibard,
Phys.\ Rev.\ Lett.\ {\bf 84}, 806 (2000);
R.~Abo-Shaeer, C.~Raman, J.~ M.~Vogels, and W.~Ketterle
Science {\bf 292}, 5516 (2001);
K. W. Madison, F. Chevy, V. Bretin, and J. Dalibard,
Phys.\ Rev.\ Lett.\ {\bf 86}, 4443 (2001);
S.~Donadello, S.~Serafini, M.~Tylutki, L.~P.~Pitaevskii, F.~Dalfovo, G.~Lamporesi, and G.~Ferrari,
Phys.\ Rev.\ Lett.\ {\bf113}, 065302 (2014).

\bibitem{Fetter:2009zz} 
  A.~L.~Fetter,
  %``Rotating trapped Bose-Einstein condensates,''
  Rev.\ Mod.\ Phys.\  {\bf 81}, 647 (2009).

\bibitem{GP}  
 E.~P.~Gross, Nuovo Cimento {\bf 20}, 454 (1961); 
 L.~P.~Pitaevskii, Zh.\ Eksp.\ Teor.\ Fiz.\ {\bf 40}, 646 (1961) 
 [Sov.\ Phys.\ JETP {\bf 13}, 451 (1961).]


\bibitem{fluctuation}
N.~R.~Cooper, N.~K.~Wilkin, and J.~M.~F.~Gunn,
Phys.\ Rev.\ Lett.\ {\bf 87}, 120405 (2001);
J.~Sinova, C.~B.~Hanna, and A.~H.~MacDonald,
Phys.\ Rev.\ Lett.\ {\bf 89}, 030403 (2002);
T.~Nakajima and M.~Ueda,
Phys.\ Rev.\ Lett.\ {\bf 91}, 140401 (2003).

\bibitem{Berges:2005yt} 
  J.~Berges and I.-O.~Stamatescu,
  %``Simulating nonequilibrium quantum fields with stochastic quantization techniques,''
  Phys.\ Rev.\ Lett.\  {\bf 95}, 202003 (2005)
  [hep-lat/0508030];
J.~Berges, S.~Borsanyi, D.~Sexty and I.-O.~Stamatescu,
  %``Lattice simulations of real-time quantum fields,''
  Phys.\ Rev.\ D {\bf 75}, 045007 (2007)
  [hep-lat/0609058];
%\bibitem{Fukushima:2014iqa} 
  K.~Fukushima and T.~Hayata,
  %``Schwinger Mechanism with Stochastic Quantization,''
  Phys.\ Lett.\ B {\bf 735}, 371 (2014)
  [arXiv:1403.4177 [hep-th]].
  %%CITATION = ARXIV:1403.4177;%%

\bibitem{Karsch:1985cb} 
  F.~Karsch and H.~W.~Wyld,
  %``Complex Langevin Simulation of the SU(3) Spin Model With Nonzero Chemical Potential,''
  Phys.\ Rev.\ Lett.\  {\bf 55}, 2242 (1985);
%\bibitem{Ambjorn:1986fz} 
  J.~Ambjorn, M.~Flensburg and C.~Peterson,
  %``The Complex Langevin Equation and Monte Carlo Simulations of Actions With Static Charges,''
  Nucl.\ Phys.\ B {\bf 275}, 375 (1986);
%\bibitem{Aarts:2008rr} 
  G.~Aarts and I.~O.~Stamatescu,
  %``Stochastic quantization at finite chemical potential,''
  JHEP {\bf 0809}, 018 (2008)
  [arXiv:0807.1597 [hep-lat]];
  G.~Aarts,
  %``Can stochastic quantization evade the sign problem? The relativistic Bose gas at finite chemical potential,''
  Phys.\ Rev.\ Lett.\  {\bf 102}, 131601 (2009)
  [arXiv:0810.2089 [hep-lat]];
  D.~Sexty,
  %``Simulating full QCD at nonzero density using the complex Langevin equation,''
  Phys.\ Lett.\ B {\bf 729}, 108 (2014)
  [arXiv:1307.7748 [hep-lat]].

\bibitem{Tsubota}
M.~Tsubota, K.~Kasamatsu, and M.~Ueda,
Phys.\ Rev.\ A {\bf 65}, 023603 (2002).

\bibitem{Andersen:2004}
 J.~O.~Andersen,
 Rev.\ Mod.\ Phys.\ {\bf 76}, 599 (2004)
 [arXiv:cond-mat/0305138].

\bibitem{D'Elia:2012tr} 
  M.~D'Elia,
  %``Lattice QCD Simulations in External Background Fields,''
  Lect.\ Notes Phys.\  {\bf 871}, 181 (2013)
  [arXiv:1209.0374 [hep-lat]].

\bibitem{Hasenfratz:1983ba} 
  P.~Hasenfratz and F.~Karsch,
  %``Chemical Potential on the Lattice,''
  Phys.\ Lett.\ B {\bf 125}, 308 (1983);
%\bibitem{Chen:2003vy} 
  J.~W.~Chen and D.~B.~Kaplan,
  %``A Lattice theory for low-energy fermions at finite chemical potential,''
  Phys.\ Rev.\ Lett.\  {\bf 92}, 257002 (2004)
  [hep-lat/0308016].
  %%CITATION = HEP-LAT/0308016;%%

\bibitem{Yamamoto:2012wy} 
  A.~Yamamoto and T.~Hatsuda,
  %``Quantum Monte Carlo simulation of three-dimensional Bose-Fermi mixtures,''
  Phys.\ Rev.\ A {\bf 86}, 043627 (2012)
  [arXiv:1209.1954 [cond-mat.quant-gas]].

\bibitem{Parisi:1980ys} 
  G.~Parisi and Y.~s.~Wu,
  %``Perturbation Theory Without Gauge Fixing,''
  Sci.\ Sin.\  {\bf 24}, 483 (1981);
  G.~Parisi,
  %``On Complex Probabilities,''
  Phys.\ Lett.\ B {\bf 131}, 393 (1983).

\bibitem{Damgaard:1987rr} 
  P.~H.~Damgaard and H.~Huffel,
  %``Stochastic Quantization,''
  Phys.\ Rept.\  {\bf 152}, 227 (1987);
%\bibitem{Namiki:1993fd} 
  M.~Namiki,
  %``Basic ideas of stochastic quantization,''
  Prog.\ Theor.\ Phys.\ Suppl.\  {\bf 111}, 1 (1993).

\bibitem{Aarts:2011zn} 
  G.~Aarts and F.~A.~James,
  %``Complex Langevin dynamics in the SU(3) spin model at nonzero chemical potential revisited,''
  JHEP {\bf 1201}, 118 (2012)
  [arXiv:1112.4655 [hep-lat]].

\bibitem{chang}
Chien-Cheng Chang, Math. Comp. {\bf 49} 180 (1987) 523-542.

\bibitem{Aarts:2013uxa} 
  G.~Aarts, L.~Bongiovanni, E.~Seiler, D.~Sexty and I.~O.~Stamatescu,
  %``Controlling complex Langevin dynamics at finite density,''
  Eur.\ Phys.\ J.\ A {\bf 49}, 89 (2013)
  [arXiv:1303.6425 [hep-lat]].

\bibitem{Nishimura:2015pba} 
  J.~Nishimura and S.~Shimasaki,
  %``New insights into the problem with a singular drift term in the complex Langevin method,''
  Phys.\ Rev.\ D {\bf 92}, 011501 (2015)
  [arXiv:1504.08359 [hep-lat]].

\bibitem{Cohen:2003kd} 
  T.~D.~Cohen,
  %``Functional integrals for QCD at nonzero chemical potential and zero density,''
  Phys.\ Rev.\ Lett.\  {\bf 91}, 222001 (2003)
  [hep-ph/0307089].

\bibitem{Yang1962}
C.~Yang, 
%``Concept of off-diagonal long-range order and the quantum phases of liquid he and of superconductors,'' 
  Rev.\ Mod.\ Phys.\ {\bf34}, 694 (1962).

\bibitem{Luscher:1986pf} 
  M.~L\"uscher,
  %``Volume Dependence of the Energy Spectrum in Massive Quantum Field Theories. 2. Scattering States,''
  Commun.\ Math.\ Phys.\  {\bf 105}, 153 (1986).
  %%CITATION = CMPHA,105,153;%%

\bibitem{Viefers}
  S.~Viefers
  %``Quantum Hall physics in rotating Bose-Einstein condensates,''
  J. Phys. Cond. Mat. {\bf 20}, 123202 (2008)
  [arXiv:0801.4856 [cond-mat]].

\bibitem{BKT}
V.~L.~Berezinskii, JETP {\bf 34} 610 (1972);
J.~M.~Kosterlitz,  and D.~J.~Thouless,
J.~Phys. C {\bf 6}, 1181 (1973).

\bibitem{BKTexperiments}
  D.~J.~Bishop, and J.~Reppy, 
  Phys.\ Rev.\ Lett.\ {\bf 40}, 1727 (1978);
  M.~R.~Beasley, J.~Mooij,  and T.~Orlando,
  Phys.\ Rev.\ Lett.\ {\bf 42}, 1165 (1979);
  D.~J.~Resnick,  J.~Garland, J.~Boyd, S.~Shoemaker,  and R.~Newrock,
  Phys.\ Rev.\ Lett.\ {\bf 47}, 1542 (1981);
  Z.~Hadzibabic,  P.~{Kr{\"u}ger},  M.~{Cheneau}, B.~{Battelier},  and  J.~{Dalibard},
  Nature {\bf 441}, 1118 (2006).

\bibitem{Yamamoto:2013zwa} 
  A.~Yamamoto and Y.~Hirono,
  %``Lattice QCD in rotating frames,''
  Phys.\ Rev.\ Lett.\  {\bf 111}, 081601 (2013)
  [arXiv:1303.6292 [hep-lat]].

\bibitem{Kibble:1976sj} 
  T.~W.~B.~Kibble,
  %``Topology of Cosmic Domains and Strings,''
  J.\ Phys.\ A {\bf 9}, 1387 (1976);
  A.~Vilenkin,
  %``Cosmic Strings and Domain Walls,''
  Phys.\ Rept.\  {\bf 121}, 263 (1985);
  M.~B.~Hindmarsh and T.~W.~B.~Kibble,
  %``Cosmic strings,''
  Rept.\ Prog.\ Phys.\  {\bf 58}, 477 (1995)
  [hep-ph/9411342].

\bibitem{tc}
T.~D.~Lee and C.~N.~Yang,
Phys.\ Rev.\ {\bf 112}, 1419 (1958);
P.~Gr\"{u}ter, D.~Ceperley, and F.~Lalo\"{e},
Phys.\ Rev.\ Lett.\ {\bf 79}, 3549 (1997);
V.~A.~Kashurnikov, N.~V.~Prokof'ev, and B.~V.~Svistunov,
Phys.\ Rev.\ Lett.\ {\bf 87}, 120402 (2001);
P.~Arnold and G.~Moore,
Phys.\ Rev.\ Lett.\ {\bf 87}, 120401 (2001).

\end{thebibliography}
\end{document}